\documentclass[floats,prl,aps,amssymb,nofootinbib,twocolumn]{revtex4-2}
\usepackage{amssymb,amsbsy,epsfig,color,graphicx}
\usepackage{color}
\usepackage{[longtable}
\usepackage{array}
\usepackage{dcolumn}   
\usepackage{cellspace}
\usepackage{mathtools}
\usepackage{amstext}
\usepackage{amssymb}
\usepackage{stmaryrd}
\usepackage{stackrel}
\usepackage{graphicx}
\usepackage{esint}
\usepackage[utf8]{inputenc}
\usepackage{blindtext}
\usepackage{booktabs}
\usepackage{enumitem}
\usepackage{bbold}
\usepackage{slashed}
\usepackage{hyperref}
\usepackage{etoolbox} 
\usepackage{lipsum} 
\usepackage[capitalize]{cleveref}
\usepackage{dsfont}
\usepackage{multirow}
\usepackage[caption=false]{subfig}

\def\be{\begin{equation}}
\def\ee{\end{equation}}
\def\bea{\begin{eqnarray}}
\def\eea{\end{eqnarray}}

\newcommand{\ba}{\begin{eqnarray}}
\newcommand{\ea}{\end{eqnarray}}

\renewcommand\[{\begin{equation}}
\renewcommand\]{\end{equation}}
\makeatletter

\appto{\appendix}{%
\@ifstar{\def\theequation@prefix{A.}}%
{}%
}
\makeatother

\hypersetup{pdfstartview=FitV,colorlinks=true,linkcolor=midblue,citecolor=midblue,filecolor=midblue,urlcolor=midblue}

\definecolor{midblue}{rgb}{0,0,0.6}

\begin{document}

\title{The Ghost and Its Secret Partner}

\author{Luca Buoninfante}
\email{luca.buoninfante@usc.es}
\affiliation{Departamento de Física de Partículas, Instituto Galego de Física de Altas Enerxías (IGFAE),  Universidade de Santiago de Compostela, Spain}

\begin{abstract}
The propagator of a ghost field acquires a pair of first-sheet complex conjugate poles above the multi-particle threshold. We show that this pole structure reveals the existence of a hidden asymptotic composite field -- the ghost's secret partner -- that remains coupled to the ghost asymptotically. This coupling induces persistent interference between the negative-norm one-particle state and the positive-norm composite-particle state. We analyze this phenomenon in four-derivative theories, in particular quadratic gravity, where the quantum dynamics prevents the spin-two ghost from appearing as an isolated asymptotic excitation. Although our results do not yet rule out observable negative probabilities, they may provide a dynamical basis for alternative amplitude prescriptions or superselection rules usually chosen or imposed ad hoc in theories with indefinite-norm~vector~spaces.
\end{abstract}

\maketitle
\newpage


\textit{\textbf{Introduction.}}--- In quantum field theory (QFT), a ghost is defined as a field whose kinetic term, and so its bare propagator, has the opposite sign to that of ordinary fields. Working in natural units ($\hbar = 1 = c$) and adopting the mostly-plus signature for the spacetime metric, the bare propagator of a scalar ghost of mass $m>0$ reads
\begin{equation}
G(-p^2)=\frac{-i\,a}{p^2+m^2-i\epsilon}\,,\qquad a=-1\,.
\label{bare-propag}
\end{equation}
The Feynman prescription $\epsilon\to0^+$ is compatible with  unitary evolution in an indefinite-norm vector space, i.e. $U^\dagger\mathbb{1}U=\mathbb{1}$, where $\mathbb{1}=\sum_{\lbrace n\rbrace}\sigma_n|n\rangle\langle n|$, with $\sigma_n=+1$ ($-1$) for states containing an even (odd) number of ghosts~\cite{Nakanishi:1972pt}.

The characteristic ``minus'' sign can improve the high-energy behavior of certain theories via enhanced cancellations of ultraviolet divergences in the loop expansion. This occurs, for example, in four-derivative QFTs such as Lee-Wick theories~\cite{Lee:1969fy,Lee:1970iw,Coleman:1969xz,Donoghue:2018lmc} and quadratic gravity~\cite{Stelle:1976gc,Tomboulis:1977jk,Salvio:2018crh,Anselmi:2018tmf,Donoghue:2021cza,Holdom:2021hlo,Buoninfante:2023ryt}. At the same time, vector spaces with negative-norm states have long raised concerns about the probabilistic interpretation of the quantum theory~\cite{Salvio:2015gsi,Kubo:2023lpz,Holdom:2024onr}.

Negative norms may not be problematic if ghosts are not asymptotic states, for instance because they are unstable and short-lived. This idea was originally proposed by Lee and Wick~\cite{Lee:1969fy,Lee:1970iw} and later revisited in~\cite{Grinstein:2008bg,Donoghue:2019fcb}. However, it has been shown that within QFT ghosts do not decay but instead remain part of the asymptotic spectrum regardless of the kinematic regime~\cite{Kubo:2024ysu}.

In this Letter, we unveil distinctive features of ghosts above the multi-particle threshold. Unlike an ordinary field, a ghost remains quadratically coupled to its composite source field in the asymptotic limit. This coupling leads to persistent interference between the negative-norm one-particle state and the positive-norm composite-particle state, precluding an interpretation of the ghost as an isolated asymptotic particle excitation.

We first illustrate the phenomenon in a two-derivative scalar model, generalizing the analysis of~\cite{Buoninfante:2026mve} to generic interactions, then extend it to four-derivative scalar theories and finally to quadratic gravity. We conclude by discussing implications and open questions.

\textit{\textbf{Ghosts and anti-instability.}}--- Let us begin by clarifying why a ghost coupled to ordinary fields cannot decay. Consider the action terms
\begin{equation}
	S\supset \int{\rm d}^4x \left[\frac{1}{2}\partial_\mu \phi(x)\partial^\mu\phi(x)+\frac{m^2}{2}\phi^2(x)\right]+S_{\rm int}[\phi]\,,
	\label{lagr-terms}
\end{equation}
where $\phi(x)$ is the ghost field and $S_{\rm int}[\phi]$ accounts for its self-interactions and interactions with ordinary fields. We assume that the ghost is the heaviest field and that the interaction term generates self-energy contributions whose resummation yields the dressed propagator~\cite{Buoninfante:2025klm}
\begin{equation}
\bar{G}(-p^2)=\frac{+i}{p^2+m^2+\Sigma(-p^2)}\,.
\label{dress-propag}
\end{equation}
The self-energy develops a nonzero imaginary part for $-p^2>M_{\rm th}^2$, where $M_{\rm th}$ is the multi-particle mass threshold. Since the sign of $\Sigma(-p^2)$ in~\eqref{dress-propag} is opposite to that in the ordinary case, the resulting complex conjugate poles lie on the first, rather than the second, Riemann sheet. 

Therefore, the first-sheet spectral representation of~\eqref{dress-propag} above the multi-particle threshold reads~\cite{Kubo:2024ysu,Buoninfante:2025klm}
\begin{eqnarray}
	\left\langle\bar{0}|{\rm T}[\phi(x)\phi(y)]|\bar{0} \right\rangle\!\!\!&=&\!\!\!-Z\Delta_{\rm F}(x\!-\!y;\!M^2)\!-\!Z^*\Delta_{\rm F}(x\!-\!y;\!M^{*2})\nonumber\\
	&&\!\!\!+\int_{M^2_{\rm th}}^\infty{\rm d}\sigma \rho(\sigma)\Delta_{\rm F}(x\!-\!y;\sigma)\,,
	\label{spectral-represent-ghost}
\end{eqnarray}
where ${\rm T}[\cdots]$ denotes the time ordering, $|\bar{0}\rangle$ the interacting vacuum,  $\Delta_{\rm F}(x-y;s)=\int_{\mathcal{C}} \frac{{\rm d}p_0}{2\pi}\int\frac{{\rm d}^3p}{(2\pi)^3} {\rm e}^{ip\cdot (x-y)} \frac{-i}{p^2+s-i\epsilon}$ are Feynman-like propagators with $s=M^2, M^{*2}, \sigma$, and $\rho(\sigma)={\rm Re}\big[\bar{G}(\sigma+i\epsilon)\big]/\pi$ is the spectral density associated with the branch-cut discontinuity. The poles are located at $M^2 \equiv m^2 + i m \Gamma$ and $M^{*2} \equiv m^2 - i m \Gamma$, where $m$ and $\Gamma$ are the renormalized mass and the width, respectively, consistently with the renormalization condition ${\rm Re}[\Sigma(M^2)]=0$ and the pole equation ${\rm Im}[\Sigma(M^2)]=m\Gamma$. The  residues are defined by $Z^{-1} = 1 - \left.{\rm d}\Sigma(z)/{\rm d}z\right|_{z=M^2}$. $\mathcal{C}$~is the Lee-Wick contour~\cite{Lee:chicago}, obtained by deforming the Feynman contour to avoid pole crossing when the real pole splits into a complex-conjugate pair due to radiative corrections. It coincides with the standard Feynman contour for $s=\sigma, M^{*2}$; see~\cite{Nakanishi:1972wx,Buoninfante:2026mve} for details.

Using the expression~\eqref{spectral-represent-ghost}, one can also obtain the spectral representation of the commutator $\langle\bar{0}|[\phi(x),\phi(y)]|\bar{0}\rangle$, which is useful for deriving the following sum rule~\cite{Kubo:2024ysu}:
\begin{eqnarray}
-1=-(Z+Z^*)+C\,,\qquad C\equiv \int_{M_{\rm th}^2}^\infty{\rm d}\sigma \rho(\sigma)\,,
\label{sum-rule-rel}
\end{eqnarray}
where $C>0$ if positive-norm multi-particle states dominate the continuum contribution to the ghost propagator, as is generally the case in the scenarios considered here.

For ordinary unstable particles, the sum rule~\eqref{sum-rule-rel} is replaced by $1=C$, indicating complete decay into multi-particle states. A ghost, instead, remains in the asymptotic spectrum: the initial norm contribution $-1$ is only partially converted into $C$, while $-(Z+Z^*)$ retains the ghost-like character, as required by unitarity (i.e. norm conservation). This behavior, distinct from both stable and unstable scenarios, was termed \textit{anti-instability}~\cite{Kubo:2024ysu}.

\textbf{\textit{Asymptotic~dynamics.}}--- The (weak) asymptotic limit of the ghost field,
\begin{eqnarray}
\phi(x)\xrightarrow[x^0\rightarrow \pm \infty]{} Z_\phi^{1/2} \phi^{\rm as}(x)\,,
\label{ghost-asymp}
\end{eqnarray}
must be such that $Z_\phi\langle 0|{\rm T}[\phi^{\rm as}(x)\phi^{\rm as}(y)]|0\rangle$ matches the pole terms of~\eqref{spectral-represent-ghost}, where $Z_\phi\in\mathbb{R}$ is the wave-function renormalization constant and $|0\rangle$ the asymptotic vacuum. Unlike ordinary asymptotic fields whose kinetic operator is diagonal, there is no local Hermitian Lagrangian for the single field $\phi^{\rm as}(x)$ that reproduces the pole part in~\eqref{spectral-represent-ghost}. A local and Hermitian description instead requires \textit{doubling} the field content and introducing quadratic couplings.

Using the field equation for the ghost field,
\begin{eqnarray}
\!\!(\Box-m^2)\phi(x)=-m\Gamma \tilde{\phi}(x)\,,\quad\,\,  \tilde{\phi}(x)\equiv \frac{-1}{m\Gamma}\frac{\delta S_{\rm int}[\phi]}{\delta \phi(x)}\,,
\label{ghost-EOM}
\end{eqnarray}
and acting with $(\Box_x-m^2)(\Box_y-m^2)$ on~\eqref{spectral-represent-ghost}, we obtain
\begin{eqnarray}
	&&\!\!\!\!\!\!\!\!\!\!\big\langle\bar{0}\big|{\rm T}\big[\tilde{\phi}(x)\tilde{\phi}(y)\big]\big|\bar{0} \big\rangle\!=\!Z\Delta_{\rm F}(x\!-\!y;\!M^2)\!+\!Z^*\Delta_{\rm F}(x\!-\!y;\!M^{*2})\nonumber\\
	&&\!\!\!\!\!\!\!\!\!\,+\int_{M^2_{\rm th}}^\infty{\rm d}\sigma \rho(\sigma)\frac{(\sigma-m^2)^2}{m^2\Gamma^2}\Delta_{\rm F}(x\!-\!y;\sigma)+g(x,y)\,,\,\,\,
	\label{spectral-represent-composite}
\end{eqnarray}
where $g(x,y)$ is a contact term whose explicit form is not important for our purposes; its expression can be found in~\cite{Buoninfante:2026mve,Buoninfante:2026jfj}. We have defined $\tilde{\phi}(x)$ as the composite field that, up to a constant factor, sources the ghost $\phi(x)$.

The pole terms in~\eqref{spectral-represent-composite} have the opposite sign to those in~\eqref{spectral-represent-ghost}. This crucial feature points to a local and Hermitian  Lagrangian reproducing both pole structures. Defining the asymptotic field $\tilde{\phi}^{\rm as}(x)$ through the (weak) limit
\begin{eqnarray}
	\tilde{\phi}(x)\xrightarrow[x^0\rightarrow \pm \infty]{} Z_{\tilde{\phi}}^{1/2} \tilde{\phi}^{\rm as}(x)\,,
	\label{composite-asymp}
\end{eqnarray}
the desired effective asymptotic Lagrangian reads~\cite{Buoninfante:2026mve}
\begin{eqnarray}
	\mathcal{L}^{\rm as}_{\phi\tilde{\phi}}\!\!\!&=&\!\!\!\frac{1}{2}\Big(\!\partial_\mu\phi^{\rm as}\partial^\mu\phi^{\rm as}\!+\!m^2\phi^{{\rm as}2}\!\Big)\!-\!\frac{1}{2}\Big(\!\partial_\mu\tilde{\phi}^{\rm as}\partial^\mu\tilde{\phi}^{\rm as}\!+\!m^2\tilde{\phi}^{{\rm as}2}\!\Big)\nonumber\\
	&&\!\!\!\!+\!\tan \theta_{\rm Z}\Big[\partial_\mu \phi^{\rm as}\partial^\mu \tilde{\phi}^{\rm as}\!+\!m^2\phi^{\rm as} \tilde{\phi}^{\rm as}\!+\!\frac{m\Gamma}{2}\!\left(\!\phi^{{\rm as}2}\!-\!\tilde{\phi}^{{\rm as}2}\!\right)\! \Big]\nonumber\\
	&&\!\!\!\!-m\Gamma \phi^{\rm as} \tilde{\phi}^{\rm as}\,,
	\label{asymp-phiphitilde-lagran}
\end{eqnarray}
where $Z_{\tilde{\phi}}=Z_{\phi}=2|Z|/\cos\theta_{\rm Z}$ was used to ensure canonical normalization, with $\theta_{\rm Z}\equiv\arctan\left({\rm Im}[Z]/{\rm Re}[Z]\right)$. Note that, while $\phi^{\rm as}(x)$ is ghost-like, $\tilde{\phi}^{\rm as}(x)$ is not.

The corresponding propagator is a symmetric, non-diagonal $2\times 2$ matrix, whose four elements in momentum space are~\cite{Buoninfante:2026mve}
\begin{eqnarray}
\!\!\!\!\!\!\!	G_{\phi^{\rm as}\phi^{\rm as}}=-G_{\tilde{\phi}^{\rm as}\tilde{\phi}^{\rm as}}\!\!&=&\!\!\frac{1}{Z_\phi}\left[\frac{iZ}{p^2+M^2}+\frac{iZ^*}{p^2+M^{*2}}\right],\,\,\,\label{propag-phiphitilde}\\
\!\!\!\!\!\!\!	G_{\phi^{\rm as} \tilde{\phi}^{\rm as}}=G_{\tilde{\phi}^{\rm as}\phi^{\rm as}}\!\!&=&\!\!\frac{1}{Z_\phi}\left[\frac{Z}{p^2+M^2}-\frac{Z^*}{p^2+M^{*2}}\right].\,\,\,
	\label{propagators-phiphitilde-mixed}
\end{eqnarray}
The diagonal elements, multiplied by $Z_\phi$, reproduce the pole terms of~\eqref{spectral-represent-ghost} and~\eqref{spectral-represent-composite}. The nonvanishing off-diagonal elements signal interference between $\phi(x)$ and $\tilde{\phi}(x)$ that persists in the asymptotic limit. Consequently, the ghost does not decouple asymptotically because of its persistent quadratic coupling to the composite source field.

The asymptotic-field equations are equivalent to the following coupled system:
\begin{eqnarray}
\!\!\!\!\!\!\!	(\Box\!-\!m^2)\phi^{\rm as}\!=\!-m\Gamma \tilde{\phi}^{\rm as}\,,\quad 
	(\Box\!-\!m^2)\tilde{\phi}^{\rm as}\!=\! m\Gamma\phi^{\rm as}\,.
	\label{EOM-phiphitilde}
\end{eqnarray}

\textbf{\textit{Asymptotic states.}}--- To find the asymptotic states, it is convenient to first diagonalize the Lagrangian~\eqref{asymp-phiphitilde-lagran}. This is achieved by introducing the complex field $\psi(x)$ and its Hermitian conjugate $\psi^\dagger(x)$ through the change of field basis
\begin{eqnarray}
\!\!\!\!\!\!\phi^{\rm as}(x)&=&Z^{-1/2}_\phi\left[Z^{1/2}\psi(x)+Z^{*1/2}\psi^\dagger(x)\right], \label{change-phi}\\
\!\!\!\!\!\!\tilde{\phi}^{\rm as}(x)&=&-i\,Z^{-1/2}_\phi\left[Z^{1/2}\psi(x)-Z^{*1/2}\psi^\dagger(x)\right],
\label{change-phitilde}
\end{eqnarray}
which leads to the equivalent Lagrangian
\begin{eqnarray}
	\!\!\!\!\!\!\mathcal{L}^{\rm as}_{\psi\psi^\dagger}\!=\!\frac{1}{2}\Big[\partial_\mu \psi \partial^\mu \psi\!+\!M^2\psi^2\!+\!\partial_\mu \psi^\dagger \partial^\mu \psi^\dagger\!+\!M^{*2}\psi^{\dagger 2}\Big]\,.
	\label{complex-lagr}
\end{eqnarray}
Defining $\Omega_{\vec{p}}\equiv\sqrt{\vec{p}^2+M^2}$ and using the decomposition
\begin{equation}
	\psi(x)\!=\!\int\!\frac{{\rm d}^3p}{\sqrt{(2\pi)^32\Omega_{\vec{p}}}}
	\left(\alpha_{\vec{p}}\,{\rm e}^{-i\Omega_{\vec{p}}x^0+i{\vec{p}\cdot\vec{x}}}\!+\!\beta^\dagger_{\vec{p}}\,{\rm e}^{i\Omega_{\vec{p}}x^0-i{\vec{p}\cdot\vec{x}}}\right),
	\label{compl-field-decomp}
\end{equation}
the system~\eqref{complex-lagr} can be canonically quantized~\cite{Nakanishi:1972wx,Kubo:2024ysu,Buoninfante:2026mve}. The asymptotic vacuum $|0\rangle$, defined by  $\alpha_{\vec{p}}|0\rangle\!=\!\beta_{\vec{p}}|0\rangle\!=\!0,$ is the state annihilated by the Hamiltonian associated with~\eqref{complex-lagr}, $H=-\int {\rm d}^3p\big[\Omega_{\vec{p}}\beta^\dagger_{\vec{p}}\alpha_{\vec{p}}+\Omega^*_{\vec{p}}\alpha^\dagger_{\vec{p}}\beta_{\vec{p}}\big]$. Using~\eqref{change-phi} and~\eqref{change-phitilde}, one readily verifies that $Z_\phi\langle 0|{\rm T}[\phi^{\rm as}(x)\phi^{\rm as}(y)]|0\rangle$ reproduces the pole structure in~\eqref{spectral-represent-ghost} and~\eqref{propag-phiphitilde}.

The ``one-particle'' Hamiltonian eigenstates are 
\begin{equation}
\big|\psi_{\vec{p}}(x^0)\big\rangle\!\equiv\! {\rm e}^{i\Omega_{\vec{p}}x^0}\!\beta_{\vec{p}}^\dagger\big|0\big\rangle \,,\quad
\big|\psi^\dagger_{\vec{p}}(x^0)\big\rangle\!\equiv\!  {\rm e}^{i\Omega^*_{\vec{p}}x^0}\!\alpha_{\vec{p}}^\dagger\big|0\big\rangle\,,
\label{psipsidagger-state}
\end{equation}
with eigenvalues $\Omega_{\vec{p}}$ and $\Omega_{\vec{p}}^*$, respectively. Since the only nonvanishing commutators are $\big[\alpha_{\vec{p}},\beta^\dagger_{\vec{k}}\big]=\big[\beta_{\vec{p}},\alpha^\dagger_{\vec{k}}\big]=-\delta^{(3)}\big(\vec{p}-\vec{k}\big)$, the states in~\eqref{psipsidagger-state} have nonvanishing equal-time mixed inner product and vanishing norm:
\begin{eqnarray}
	\!\!\!\!\!\!\!\!\!\!\!\!\!\!\!\! &&\big\langle\psi_{\vec{p}}(x^0)\big|\psi^\dagger_{\vec{k}}(x^0)\big\rangle\!=\!\big\langle\psi^\dagger_{\vec{p}}(x^0)\big|\psi_{\vec{k}}(x^0)\big\rangle\!=\!-\delta^{(3)}\big(\vec{p}-\vec{k}\big),\nonumber\\
	\!\!\!\!\!\!\!\!\!\!\!\!\!\!\!\! &&\big\langle\psi_{\vec{p}}(x^0)\big|\psi_{\vec{k}}(x^0)\big\rangle\!=\!\big\langle\psi^\dagger_{\vec{p}}(x^0)\big|\psi^\dagger_{\vec{k}}(x^0)\big\rangle\!=\!0\,.
	\label{zero-norm}
\end{eqnarray}
Consequently, the asymptotic states associated with $\phi(x)$ and $\tilde{\phi}(x)$ follow from the change of basis in~\eqref{change-phi} and~\eqref{change-phitilde}. Up to overall constant factors, they are given by
\begin{eqnarray}
	&&\!\!\!\!\!\!\!\!\!\!\!\!\!\!\!\!\big|\phi^{\rm as}_{\vec{p}}(x^0)\big\rangle\!\equiv\! \frac{1}{\sqrt{\cos\theta_{\rm Z}}}\int\!\! {\rm d}^3x \sqrt{\!\frac{2|\Omega_{\vec{p}}|}{(2\pi)^3}} {\rm e}^{i\vec{p}\cdot \vec{x}}\phi^{\rm as}(x) \big|0\big\rangle\label{phi-state}\\
	&&\!\!\!\!\!\!\!\!\!\!\!\!=\! \frac{1}{\sqrt{2}}\!\left[{\rm e}^{i(\theta_{\vec{p}}-\theta_{\rm Z})/2}\big|\psi^\dagger_{\vec{p}}(x^0)\big\rangle+{\rm e}^{-i(\theta_{\vec{p}}-\theta_{\rm Z})/2}\big|\psi_{\vec{p}}(x^0)\big\rangle\right], \nonumber  \\
	&&\!\!\!\!\!\!\!\!\!\!\!\!\!\!\!\!\big|\tilde{\phi}^{\rm as}_{\vec{p}}(x^0)\big\rangle\!\equiv\! \frac{1}{\sqrt{\cos\theta_{\rm Z}}}\int\!\! {\rm d}^3x \sqrt{\!\frac{2|\Omega_{\vec{p}}|}{(2\pi)^3}} {\rm e}^{i\vec{p}\cdot \vec{x}}\tilde{\phi}^{\rm as}(x) \big|0\big\rangle\label{phitilde-state}\\
	&&\!\!\!\!\!\!\!\!\!\!\!\!=\!\frac{i}{\sqrt{2}}\!\left[{\rm e}^{i(\theta_{\vec{p}}-\theta_{\rm Z})/2}\big|\psi^\dagger_{\vec{p}}(x^0)\big\rangle-{\rm e}^{-i(\theta_{\vec{p}}-\theta_{\rm Z})/2}\big|\psi_{\vec{p}}(x^0)\big\rangle\right], \nonumber 
\end{eqnarray}
where $\theta_{\vec{p}}\equiv \arctan ({\rm Im}[\Omega_{\vec{p}}]/{\rm Re}[\Omega_{\vec{p}}]).$ These states are non-orthogonal and have negative and positive norms, respectively, given by $\mp \delta^{(3)}(\vec{p}-\vec{k})\cos(\theta_{\vec{p}}-\theta_{\rm Z})$. Consistency requires $\cos(\theta_{\vec{p}}-\theta_{\rm Z})>0$, or equivalently ${\rm Re}[\Omega_{\vec{p}}Z^*]>0$. We refer to~\eqref{phitilde-state} as a composite-particle state, despite the fact that it is not an ordinary bound state.

\textbf{\textit{Persistent interference.}}--- The physical picture that emerges is the following. Above the multi-particle threshold, the quantum dynamics generates first-sheet complex conjugate poles that reveal the presence of a hidden asymptotic field --~the ghost's secret partner~-- which remains persistently coupled to the ghost, as described by the Lagrangian~\eqref{asymp-phiphitilde-lagran}. This coupling induces interference between the negative-norm one-particle state~\eqref{phi-state} and the positive-norm composite-particle state~\eqref{phitilde-state}, thus preventing the existence of isolated ghost excitations and precluding a particle interpretation. Although this phenomenon does not yet rule out observable negative probabilities, it is tempting to interpret it as a \textit{masking} of negative norms by the positive-norm multi-particle sector, since the free asymptotic~states~\eqref{psipsidagger-state}~have~zero~norm.

A simple model is given by $S_{\rm int}[\phi]\propto \int{\rm d}^4 x\phi(x)\chi^2(x)$, where $\chi(x)$ is an elementary scalar field of mass $\mu<m/2$. The asymptotic fields are then $\phi^{\rm as}(x)$, $\tilde{\phi}^{\rm as}(x)\propto [\chi^2]^{\rm as}(x)$, and $\chi^{\rm as}(x)$. While the latter satisfies the free Klein-Gordon equation, the other two are coupled as in~\eqref{EOM-phiphitilde}. In this model,~\eqref{phitilde-state} may be viewed as a state composed of two $\chi$~particles~\cite{Buoninfante:2026mve}. It is worth noting that, if $\phi(x)$ were an ordinary field, its one-particle state would decay and $\tilde{\phi}(x)$ would have no associated asymptotic field either, leaving only~$\chi^{\rm as}(x)$ in the asymptotic spectrum~\cite{Buoninfante:2026jfj}.

\textbf{\textit{Four-derivative theories.}}--- More relevant examples are fourth-order theories whose bare propagators contain an ordinary massless pole and a massive ghost~pole:
\begin{equation}
	S\supset\int{\rm d}^4x \left[-\frac{1}{2}\partial_\mu \Phi\left(1-\frac{\Box}{m^2}\right)\partial^\mu\Phi\right]+S_{\rm int}[\Phi],
	\label{four-der-lagr-terms}
\end{equation}
where $\Phi(x)$ is a scalar field and $S_{\rm int}[\Phi]$ accounts for its self-interactions and interactions with ordinary fields. In this case, the dressed propagator is given by~\cite{Donoghue:2018lmc,Donoghue:2019fcb,Buoninfante:2025klm}
\begin{equation}
\bar{G}_4(-p^2)=\frac{-i}{p^2+(p^2)^2/m^2-\Sigma_4(-p^2)}\,.
\label{dress-propag-fourth}
\end{equation}
We are interested in scenarios in which radiative corrections leave the massless pole unshifted; thus, we assume $\Sigma_4(-p^2)=(-p^2/m^2)\Sigma(-p^2)$, with $\Sigma(-p^2)$ playing the  role as the self-energy in~\eqref{dress-propag}. The renormalization condition ${\rm d}\Sigma_4(z)/{\rm d}z|_{z=0}=0$ fixes the residue at $-p^2=0$~to~$i$.

The spectral representation associated with~\eqref{dress-propag-fourth} reads
\begin{eqnarray}
&&\!\!\!\!\!\!\!\!\!\!\!\!\!\!\!\!\!\left\langle\bar{0}|{\rm T}[\Phi(x)\Phi(y)]|\bar{0} \right\rangle=\Delta_{\rm F}(x-y;0)-Z\Delta_{\rm F}(x\!-\!y;M^2) \nonumber\\
&&\!\!\!\!\!\!\!\!\!\!\!\!-Z^*\Delta_{\rm F}(x\!-\!y;M^{*2})+\int_{M^2_{\rm th}}^\infty{\rm d}\sigma \rho(\sigma)\Delta_{\rm F}(x\!-\!y;\sigma)\,,
	\label{spectral-represent-fourth}
\end{eqnarray}
where $Z$ and $\rho(\sigma)$ are defined as in the two-derivative case, up to factors of $m^2/M^2$ and $m^2/\sigma$, respectively; we keep the same notation for simplicity. The same sum rule~\eqref{sum-rule-rel} holds~\cite{Coleman:1969xz}, but is better understood as $1-1=1-(Z+Z^*)+C$, where $1=1$ corresponds to the massless pole and $-1=-(Z+Z^*)+C$ to the massive ghost pole.

Using the field equation for $\Phi(x),$ 
\begin{equation}
\Box\left(1-\frac{\Box}{m^2}\right)\Phi(x)=m\Gamma \tilde{\Phi}(x)\,,\quad  \tilde{\Phi}(x)\equiv \frac{-1}{m\Gamma}\frac{\delta S_{\rm int}[\Phi]}{\delta \Phi(x)}\,,
\label{four-der-EOM}
\end{equation}
and acting with $\Box_x(1-\Box_x/m^2)\Box_y(1-\Box_y/m^2)$ on~\eqref{spectral-represent-fourth}, we obtain the spectral representation for the propagator of the composite field $\tilde{\Phi}(x):$
\begin{eqnarray}
	&&\!\!\!\!\!\!\!\!\!\!\!\!\!\big\langle\bar{0}\big|{\rm T}\big[\tilde{\Phi}(x)\tilde{\Phi}(y)\big]\big|\bar{0} \big\rangle\!=\nonumber\\
    &&\!\!\!\!\!\!\!\!\!\!\!\!\!=Z\frac{M^4}{m^4}\Delta_{\rm F}(x\!-\!y;M^2)+Z^*\frac{M^{*4}}{m^4}\Delta_{\rm F}(x\!-\!y;M^{*2})	\label{spectral-represent-composite-four}\\
	&&\!\!\!\!\!\!\!+\int_{M^2_{\rm th}}^\infty{\rm d}\sigma \rho(\sigma)\frac{\sigma^2}{m^4}\frac{(\sigma-m^2)^2}{m^2\Gamma^2}\Delta_{\rm F}(x\!-\!y;\sigma)+g_4(x,y)\,,\nonumber
\end{eqnarray}
where $g_4(x,y)$ accounts for contact terms.

The propagator~\eqref{spectral-represent-composite-four} has the same complex conjugate poles as~\eqref{spectral-represent-fourth}. However, unlike in the simpler two-derivative model analyzed above, their residues differ not only in sign but also by factors of $M^4/m^4$ and $M^{*4}/m^4$. This means that the asymptotic limit of $\tilde{\Phi}(x)$ cannot simply be identified with $\tilde{\phi}^{\rm as}(x)$. Indeed, the asymptotic fields $\Phi^{\rm as}(x)$ and $\tilde{\Phi}^{\rm as}(x)$ are given by
\begin{eqnarray}
\!\!\!\!\!\!\!\!\!\!\Phi(x)\!&\xrightarrow[x^0\rightarrow \pm \infty]{}&\! \Phi^{\rm as}(x)\equiv h^{\rm as}(x)+ Z_\phi^{1/2} \phi^{\rm as}(x)\,,\\
\!\!\!\!\!\!\!\!\!\!\tilde{\Phi}(x)\!&\xrightarrow[x^0\rightarrow \pm \infty]{}&\! \tilde{\Phi}^{\rm as}(x)\equiv Z_\phi^{1/2} \left[\tilde{\phi}^{\rm as}(x)+\frac{\Gamma}{m}\phi^{\rm as}(x)\right],
\end{eqnarray}
where $h^{\rm as}(x)$ is an ordinary asymptotic field satisfying the free massless Klein-Gordon equation, while $\phi^{\rm as}(x)$ and $\tilde{\phi}^{\rm as}(x)$ are the ghost and its partner, respectively, whose dynamics is governed by~\eqref{asymp-phiphitilde-lagran}. We can approximate $\tilde{\Phi}^{\rm as}(x)\simeq Z_{\phi}^{\rm 1/2}\tilde{\phi}^{\rm as}(x)$ only if $\Gamma/m\ll 1$.

One can verify that the field equations~\eqref{EOM-phiphitilde} are consistent with $\Box(1-\Box/m^2)\Phi^{\rm as}=m\Gamma \tilde{\Phi}^{\rm as}$ and that, using $\langle 0|{\rm T}[h^{\rm as}(x)h^{\rm as}(y)]|0\rangle=\Delta_{\rm F}(x-y;0)$, the asymptotic-vacuum expectation values $\langle 0|{\rm T}[\Phi^{\rm as}(x)\Phi^{\rm as}(y)]|0\rangle$ and $\langle 0|{\rm T}[\tilde{\Phi}^{\rm as}(x)\tilde{\Phi}^{\rm as}(y)]|0\rangle$ reproduce the pole terms in~\eqref{spectral-represent-fourth} and~\eqref{spectral-represent-composite-four}, respectively, including the correct residues.

It is now clear that the physical implications discussed in the two-derivative case extend directly to four-derivative theories: the negative-norm ghost cannot be interpreted as an isolated asymptotic particle excitation.

\textbf{\textit{Quadratic gravity.}}--- A physically relevant four-derivative theory is quadratic gravity~\cite{Stelle:1976gc,Tomboulis:1977jk,Salvio:2018crh,Anselmi:2018tmf,Donoghue:2021cza,Holdom:2021hlo,Buoninfante:2023ryt}, whose classical action, up to boundary terms, reads
\begin{equation}
S[g]\!=\!\frac{M^2_{\rm Pl}}{2}\int{\rm d}^4x\sqrt{-g}\!\left[R\!+\!\frac{1}{6m_0^2}R^2\!-\!\frac{1}{2m_2^2}C_{\mu\nu\rho\sigma}C^{\mu\nu\rho\sigma}\right],
\label{lagr-quad-grav}
\end{equation}
where $C_{\mu\nu\rho\sigma}$ is the Weyl tensor, $M_{\rm Pl}$  the reduced Planck mass, and $m_{0,2}$ the masses of the ordinary spin-zero field (the \textit{scalaron}) and the spin-two ghost field, respectively. Here, we assume a negligible~cosmological~constant.

Expanding around the Minkowski background as $g_{\mu\nu}(x)=\eta_{\mu\nu}+\frac{2}{M_{\rm Pl}}H_{\mu\nu}(x)$, the total action for the gravitational field $H_{\mu\nu}(x)$ takes the schematic form~\cite{Basile:2024oms}
\begin{equation}
S[H]=\frac{1}{2}\int {\rm d}^4x H_{\mu\nu} \mathbb{K}^{\mu\nu\rho\sigma}H_{\rho\sigma} +S_{\rm int}[H]\,,
\label{action-H-total}
\end{equation}
where gauge-fixing contributions are also assumed to be included, $\mathbb{K}^{\mu\nu\rho\sigma}$ is the kinetic operator, and $S_{\rm int}[H]$ accounts for both self-interactions and matter couplings. Since the spin-two and spin-zero sectors decouple at quadratic order in $H_{\mu\nu}$, we can just focus on the spin-two components of the propagators in what follows.

Introducing the spin-two projector operator in coordinate space~\cite{Basile:2024oms}, $\mathcal{P}^{(2)}_{\mu\nu\rho\sigma}=(\theta_{\mu\rho}\theta_{\nu\sigma}+\theta_{\mu\sigma}\theta_{\nu\rho})/2-\theta_{\mu\nu}\theta_{\rho\sigma}/3$, with $\theta_{\mu\nu}=\eta_{\mu\nu}-\partial_\mu \partial_\nu/\Box$, we define the traceless and transverse spin-two component of the gravitational field as $H^{(2)}_{\mu\nu}\equiv \mathcal{P}^{(2)\rho\sigma}_{\mu\nu}H_{\rho\sigma}$. Its field equations then read
\begin{eqnarray}
\Box\left(1-\frac{\Box}{m_2^2}\right)H^{(2)}_{\mu\nu}(x)=m_2\Gamma \tilde{H}^{(2)}_{\mu\nu}(x)\,,\label{EOM-spin-two-Hmunu}
\end{eqnarray}
where we have also defined the composite spin-two field
\begin{eqnarray}
\tilde{H}^{(2)}_{\mu\nu}(x)\equiv \frac{-1}{m_2\Gamma}\mathcal{P}^{(2)\rho\sigma}_{\mu\nu}\frac{\delta S_{\rm int}[H]}{\delta H^{\rho\sigma}(x)}\,,
\label{spin-two-composite}
\end{eqnarray}
which includes terms coming from the stress-energy tensor, higher-order matter couplings, and gravitational interactions, all projected along the spin-two component.

The spin-two sector can acquire a width through self-energy contributions from both gravitational and matter couplings. The corresponding dressed propagator is analogous to~\eqref{dress-propag-fourth}, with an additional spin-two projector~\cite{Tomboulis:1977jk,Donoghue:2018izj,Anselmi:2018tmf,Donoghue:2019fcb}. Likewise, the spectral representation is analogous to~\eqref{spectral-represent-fourth} and can be written concisely as
\begin{equation}
\big\langle\bar{0}\big|{\rm T}\big[H^{(2)}_{\mu\nu}(x)H^{(2)}_{\rho\sigma}(y)\big]\big|\bar{0} \big\rangle=\mathcal{P}^{(2)}_{\mu\nu\rho\sigma}\big\langle\bar{0}\big|{\rm T}\big[\Phi(x)\Phi(y)\big]\big|\bar{0} \big\rangle\,.
\label{spectral-represent-spin-two}
\end{equation}
The idempotency of $\mathcal{P}^{(2)}_{\mu\nu\rho\sigma}$ and its orthogonality to the other spin projectors prevent gauge-fixing terms from appearing. Moreover, the locations of the complex conjugate poles are gauge-independent~\cite{Johnston:1987ue}, while diffeomorphism invariance ensures that the massless graviton pole remains unshifted.

Acting with $\Box_x(1-\Box_x/m^2_2)\Box_y(1-\Box_y/m^2_2)$ on~\eqref{spectral-represent-spin-two}, we obtain the spectral representation for the propagator of the composite field $\tilde{H}^{(2)}_{\mu\nu}(x)$, which is analogous to~\eqref{spectral-represent-composite-four} and can therefore be written concisely as
\begin{equation}
\big\langle\bar{0}\big|{\rm T}\big[\tilde{H}^{(2)}_{\mu\nu}(x)\tilde{H}^{(2)}_{\rho\sigma}(y)\big]\big|\bar{0} \big\rangle=\mathcal{P}^{(2)}_{\mu\nu\rho\sigma}\big\langle\bar{0}\big|{\rm T}\big[\tilde{\Phi}(x)\tilde{\Phi}(y)\big]\big|\bar{0} \big\rangle\,.
\label{spectral-represent-spin-two-composite}
\end{equation}
In both~\eqref{spectral-represent-spin-two} and~\eqref{spectral-represent-spin-two-composite}, all quantities appearing in the scalar propagators have to be reinterpreted in the context of quadratic gravity; in particular, the replacements $M^2\rightarrow M_2^2\equiv m_2^2+im_2\Gamma$ and $M^{*2}\rightarrow M_2^{*2}\equiv m_2^2-im_2\Gamma$ are implicit. It is also worth clarifying that the action of the differential operator $\mathcal{P}^{(2)}_{\mu\nu\rho\sigma}$ has to be understood in momentum space upon acting on the Fourier~modes~${\rm e}^{ip\cdot x}$.

Taking into account the additional tensor structure, we can again reveal the presence of a hidden asymptotic field that now remains persistently coupled to the spin-two ghost. The asymptotic fields $H^{{\rm as}(2)}_{\mu\nu}(x)$ and $\tilde{H}^{{\rm as}(2)}_{\mu\nu}(x)$ are given by the (weak) limits
\begin{eqnarray}
\!\!\!\!\!\!\!\!\!\!\!H^{(2)}_{\mu\nu}\!&\xrightarrow[x^0\rightarrow \pm \infty]{}&\! H^{{\rm as}(2)}_{\mu\nu}\equiv h^{{\rm as}(2)}_{\mu\nu}+ Z_\phi^{1/2} \phi^{{\rm as}(2)}_{\mu\nu}\,,\\
\!\!\!\!\!\!\!\!\!\!\!\tilde{H}^{(2)}_{\mu\nu}\!&\xrightarrow[x^0\rightarrow \pm \infty]{}&\! \tilde{H}^{{\rm as}(2)}_{\mu\nu}\equiv Z_\phi^{1/2} \left[\tilde{\phi}^{{\rm as}(2)}_{\mu\nu}+\frac{\Gamma}{m_2}\phi^{{\rm as}(2)}_{\mu\nu}\right],
\end{eqnarray}
where $h^{{\rm as}(2)}_{\mu\nu}$ is the spin-two part of the asymptotic massless graviton field, satisfying the projected linearized vacuum Einstein equations, i.e. $\Box h^{{\rm as}(2)}_{\mu\nu}=0$, while $\phi^{{\rm as}(2)}_{\mu\nu}\equiv \mathcal{P}^{(2)\rho\sigma}_{\mu\nu}\phi^{\rm as}_{\rho\sigma}$ and $\tilde{\phi}^{{\rm as}(2)}_{\mu\nu}\equiv \mathcal{P}^{(2)\rho\sigma}_{\mu\nu}\tilde{\phi}^{\rm as}_{\rho\sigma}$ are the spin-two projections of the asymptotic spin-two massive ghost and its partner field, respectively. Their dynamics~is~governed~by
\begin{eqnarray}
	\mathcal{L}^{\rm as}_{\phi\tilde{\phi}}\!\!&=&\!\!-\frac{1}{2}\phi^{\rm as}_{\mu\nu}(\mathbb{E}^{\mu\nu\rho\sigma}\!-\!m^2_2\mathbb{M}^{\mu\nu\rho\sigma})\phi^{\rm as}_{\rho\sigma}\nonumber\\
    &&\!\! +\frac{1}{2}\tilde{\phi}^{\rm as}_{\mu\nu}(\mathbb{E}^{\mu\nu\rho\sigma}\!-\!m^2_2\mathbb{M}^{\mu\nu\rho\sigma})\tilde{\phi}^{\rm as}_{\rho\sigma} \!-\!m_2\Gamma\, \phi^{\rm as}_{\mu\nu}\mathbb{M}^{\mu\nu\rho\sigma}\tilde{\phi}^{\rm as}_{\rho\sigma} \nonumber\\
	&&\!\!-\tan \theta_{\rm Z}\Big[\phi^{\rm as}_{\mu\nu}(\mathbb{E}^{\mu\nu\rho\sigma}\!-\!m^2_2\mathbb{M}^{\mu\nu\rho\sigma})\tilde{\phi}^{\rm as}_{\rho\sigma} \nonumber\\
	&&\!\qquad\!-\frac{m_2\Gamma}{2}\big(\phi^{\rm as}_{\mu\nu}\mathbb{M}^{\mu\nu\rho\sigma}\phi^{\rm as}_{\rho\sigma}-\tilde{\phi}^{\rm as}_{\mu\nu}\mathbb{M}^{\mu\nu\rho\sigma}\tilde{\phi}^{\rm as}_{\rho\sigma} \big)\Big]\,,
	\label{asymp-fftilde-lagran}
\end{eqnarray}
where 
\begin{eqnarray}
\mathbb{E}^{\mu\nu\rho\sigma}\!&\equiv&\! \mathbb{M}^{\mu\nu\rho\sigma}\Box+\eta^{\mu\nu}\partial^{\rho}\partial^\sigma +\eta^{\rho\sigma}\partial^{\mu}\partial^\nu -\frac{1}{2}(\eta^{\mu\rho}\partial^{\nu}\partial^\sigma\nonumber\\
&&\!+\eta^{\mu\sigma}\partial^{\nu}\partial^\rho +\eta^{\nu\rho}\partial^{\mu}\partial^\sigma+\eta^{\nu\sigma}\partial^{\mu}\partial^\rho)\,,\\
\mathbb{M}^{\mu\nu\rho\sigma}\!&\equiv&\! \frac{1}{2}(\eta^{\mu\rho}\eta^{\nu\sigma}\!+\!\eta^{\mu\sigma}\eta^{\nu\rho})-\eta^{\mu\nu}\eta^{\rho\sigma}\,.
\label{kineti-mass-operators}
\end{eqnarray}

The asymptotic-field equations corresponding to~\eqref{asymp-fftilde-lagran} are equivalent to the following coupled system:
\begin{eqnarray}
\!\!\!\!\!\!\!\!\!\!(\mathbb{E}^{\phantom{\mu\nu}\rho\sigma}_{\mu\nu}-m^2_2\,\mathbb{M}^{\phantom{\mu\nu}\rho\sigma}_{\mu\nu})\phi^{\rm as}_{\rho\sigma}(x)\!&=&\!-m_2\Gamma\, \mathbb{M}^{\phantom{\mu\nu}\rho\sigma}_{\mu\nu}\tilde{\phi}^{\rm as}_{\rho\sigma}(x),\label{EOM-spintwo-ghost}\\
\!\!\!\!\!\!\!\!\!\!(\mathbb{E}^{\phantom{\mu\nu}\rho\sigma}_{\mu\nu}-m^2_2\,\mathbb{M}^{\phantom{\mu\nu}\rho\sigma}_{\mu\nu})\tilde{\phi}^{\rm as}_{\rho\sigma}(x)\!&=&\!m_2\Gamma\, \mathbb{M}^{\phantom{\mu\nu}\rho\sigma}_{\mu\nu}\phi^{\rm as}_{\rho\sigma}(x),
\label{EOM-spintwo-partner}
\end{eqnarray}
from which it follows that both spin-two fields are traceless and transverse, namely $\phi^{\rm as}\equiv\eta^{\mu\nu}\phi^{\rm as}_{\mu\nu}=0$, $\tilde{\phi}^{\rm as}\equiv\eta^{\mu\nu}\tilde{\phi}^{\rm as}_{\mu\nu}=0$, $\partial^\mu \phi^{\rm as}_{\mu\nu}=0$, and $\partial^\mu \tilde{\phi}^{\rm as}_{\mu\nu}=0$.

Inverting the kinetic operator in~\eqref{asymp-fftilde-lagran}, we find that the propagator projected onto the spin-two sector in momentum space is given by the matrix elements~\eqref{propag-phiphitilde} and~\eqref{propagators-phiphitilde-mixed}, multiplied by the momentum-space form of the projector $\mathcal{P}^{(2)}_{\mu\nu\rho\sigma}$. Consequently, we can verify that the asymptotic-vacuum expectation values $\langle 0|{\rm T}[H^{{\rm as} (2)}_{\mu\nu}(x)H^{{\rm as} (2)}_{\rho\sigma}(y)]|0\rangle$ and $\langle 0|{\rm T}[\tilde{H}^{{\rm as} (2)}_{\mu\nu}(x)\tilde{H}^{{\rm as} (2)}_{\rho\sigma}(y)]|0\rangle$ reproduce the pole terms in~\eqref{spectral-represent-spin-two} and~\eqref{spectral-represent-spin-two-composite}, respectively. Here, the asymptotic vacuum $|0\rangle$ is defined as the state annihilated by the Hamiltonian associated with~\eqref{asymp-fftilde-lagran}.

Hence, the asymptotic spectrum of quadratic gravity includes the massless graviton and, if stable, the scalaron, both satisfying ordinary free-field equations, as well as the coupled system consisting of the spin-two ghost and its non-ghost partner field. Unlike the former two, the latter cannot appear as isolated asymptotic excitations because of their non-diagonal quadratic dynamics, precluding a particle interpretation.

For completeness, we mention that the complex spin-two fields $\psi_{\mu\nu}(x)$ and $\psi^\dagger_{\mu\nu}(x)$ can be introduced through transformations analogous to~\eqref{change-phi} and~\eqref{change-phitilde}, allowing us to diagonalize the effective Lagrangian~\eqref{asymp-fftilde-lagran} as $-\frac{1}{2}[\psi_{\mu\nu}(\mathbb{E}^{\mu\nu\rho\sigma}\!-\!M^2_2\mathbb{M}^{\mu\nu\rho\sigma})\psi_{\rho\sigma}+\text{h.\,c.}]$, and identify the corresponding zero-norm~asymptotic~states.

\textbf{\textit{Spin-two width.}}---  The leading contributions to the spin-two width at the peak $p^2=-m_2^2$ arise from the scalaron-pair and matter-pair channels, provided $m_2>2m_0$ and $m_2>2m_\Psi$, respectively, where $m_\Psi$ denotes the mass of a matter field. The ghost-graviton-graviton channel, instead, contributes at higher-loop order and is therefore perturbatively suppressed~\cite{Anselmi:2018tmf,Holdom:2021hlo,Buoninfante:2023ryt}.

If matter couplings are absent or negligible, $m_2>2m_0$ may be desirable for a less suppressed spin-two width in a pure gravitational scenario. In this case, one-loop computations give $\Gamma\simeq \frac{m_2^3}{960\pi M^2_{\rm Pl}}(1-4m_0^2/m_2^2)^{5/2}$~\cite{Anselmi:2018tmf}.

It is worth noting that this inequality has interesting cosmological implications if we assume $R^2$-driven inflation, which requires $m_0\simeq 3\times 10^{13}\mathrm{GeV}$ to fit the CMB anisotropies~\cite{Buoninfante:2025dgy}. Indeed, given the leading-order slow-roll expression for the tensor-to-scalar ratio in quadratic gravity, $r=(24/N_{\rm e}^2)m_2^2/(m_0^2+2m_2^2)$~\cite{Salvio:2022mld,Anselmi:2020lpp,Kubo:2025jla}, we find $12/N_{\rm e}^2>r> 32/(3N_{\rm e}^2)$ for $m_2>2m_0$. In particular, taking $N_{\rm e}\simeq 60$ e-folds we get $r\sim \mathcal{O}(10^{-3})$, which is the order of magnitude that will be probed by upcoming satellite missions~\cite{LiteBIRD:2026gdf}. While this is not a prediction but rather our expectation, it is intriguing that future observations could constrain $m_2$ and, consequently, the leading channels contributing to the spin-two width.

\textbf{\textit{Open questions.}}--- The canonical formalism of QFT has allowed us to shed new light on the quantum dynamics of ghosts and reveal the presence of a hidden asymptotic field -- the ghost's secret partner -- that remains permanently coupled to the ghost, preventing the appearance of isolated negative-norm excitations in the asymptotic limit. However, several questions remain open.

This persistent quadratic coupling and the vanishing norm of the free asymptotic states do not yet, by themselves, guarantee the absence of observable negative probabilities of localization. Nevertheless, our results may provide a dynamical basis for the need for alternative amplitude prescriptions~\cite{Anselmi:2025uzj}, or new inner products~\cite{Salvio:2015gsi,Holdom:2024onr,Bateman:2026eyj}, or superselection rules~\cite{Holdom:2026qez}, usually adopted or imposed ad hoc in theories with indefinite-norm vector spaces to ensure that only positive-norm states appear in the asymptotic spectrum.

In relation to the previous point, it would also be interesting to understand whether symmetry constraints could restrict the admissible asymptotic states. In this respect, it is worth noting that~\eqref{asymp-phiphitilde-lagran} and~\eqref{complex-lagr} possess an enlarged spacetime symmetry corresponding to a ``complexified'' Poincaré algebra: $\delta \psi = (\xi^\mu-i\zeta^\mu)\partial_\mu \psi$, where $\xi^\mu=\epsilon^\mu + \omega^\mu_{\phantom{\mu}\nu} x^\nu$ (with $\omega_{\mu\nu}=-\omega_{\nu\mu})$ are associated with the ten Poincaré generators, while $\zeta^\mu=\varepsilon^\mu + \varpi^\mu_{\phantom{\mu}\nu} x^\nu$ (with $\varpi_{\mu\nu}=-\varpi_{\nu\mu})$ are associated with ten additional generators. The physical implications of this enhanced spacetime symmetry will be presented elsewhere.

Negative-norm ghosts are compatible with causal propagation: positive (negative) energies propagate forward (backward) in time~\cite{Buoninfante:2025klm,Buoninfante:2026jfj}. However, it remains unclear whether first-sheet complex conjugate poles can induce (micro)causality violations, for instance through off-shell effects from loop contributions~\cite{Grinstein:2008bg}. If so, it would be interesting to investigate whether such potential acausalities are masked by interference with the partner field.

We studied the asymptotic dynamics of the spin-two ghost perturbatively around flat spacetime. Further work is needed to understand whether and how our results extend to curved backgrounds, especially to spacetimes that are not asymptotically flat.

Ghosts are often thought to cause classical instabilities under certain assumptions~\cite{Woodard:2015zca}. However, recent studies have shown that the non-linear time evolution of some ghostly theories~\cite{Deffayet:2021nnt,ErrastiDiez:2024hfq,Deffayet:2025lnj}, including quadratic gravity~\cite{Held:2023aap,Held:2025ckb}, can be stable. It is important to understand whether and how the stability of these classical solutions relate to the asymptotic quantum dynamics discussed here, given the essential role of non-linear interactions in both cases.

Addressing these questions could provide further evidence for the viability of theories containing ghost-like degrees of freedom and negative-norm states, such as four-derivative theories, while unveiling new phenomena within the QFT framework and opening new perspectives on quantum gravity in the sub-Planckian regime~\cite{Buoninfante:2025dgy,Buoninfante:2024yth}.

\medskip 

\textbf{\textit{Acknowledgments.}}--- I am grateful to Gia Dvali and Luca Smaldone for useful discussions. I acknowledge financial support from the Xunta de Galicia (CIGUS Network), the EU through the Galicia Feder 2021-2027 Program, and the Grant CEX2023-001318-M funded by MICIU/AEI/10.13039/501100011033.


\bibliographystyle{utphys}
\bibliography{References}


\end{document}